\documentclass{aa}
\usepackage{graphicx}
\usepackage{txfonts}  
\newcommand{\iso}[2]{\mbox{$^{#1}{\rm #2}$}}
\newcommand{\Teff}{T_{\rm eff}}

\newcommand{\Vmic}{V_{\rm mic}}
\newcommand{\Vmac}{V_{\rm mac}}
\newcommand{\eps}[1]{\log\varepsilon_{\rm #1}}
\newcommand{\kms}{km\,s$^{-1}$}

\begin{document}

\title{Barium even-to-odd isotope abundance ratios in thick disk and thin disk stars
\thanks{Based on observations collected at the German Spanish Astronomical
Center, Calar Alto, Spain}}

\author{L. Mashonkina \inst{1,2,3}
\and G. Zhao \inst{3}}

\institute{
Institut f\"ur Astronomie und Astrophysik der Universit\"at
M\"unchen, Scheinerstr. 1, 81679 M\"unchen, Germany \\
\and
Institute of Astronomy, Russian Academy of Science, Pyatnitskaya 48,
119017 Moscow, Russia \\
\and  National Astronomical Observatories, Chinese Academy of
Science, A20 Datun Road, Chaoyang District, Beijing 100012, China}

\date{Received  / Accepted }

\offprints{L. Mashonkina} \mail{lyuda@usm.lmu.de and lima@inasan.ru}
\abstract{We present the Ba even-to-odd isotope abundance ratios
in 25 cool dwarf stars with the metallicity [Fe/H] ranging between 0.25
and --1.35. Our method takes advantage of the hyperfine structure
(HFS) affecting the \ion{Ba}{ii} resonance line of the odd
isotopes. The fractional abundance of the odd isotopes of Ba is
derived from a requirement that Ba abundances from the resonance
line $\lambda\,4554$ and subordinate lines $\lambda\,5853$ and
$\lambda\,6496$ must be equal. The results are based on NLTE line
formation and analysis of high resolution (R $\sim60000$) high
signal-to-noise (S/N $\ge 200$) observed spectra. We find that the
fraction of the odd isotopes of Ba grows toward the lower Ba
abundance (or metallicity) and the mean value in the thick disk
stars equals 33 $\pm$ 4\%. This indicates the higher contribution
of the $r-$process to barium in the thick disk stars compared to
the solar system matter. The obtained fraction increases with the
[Eu/Ba] abundance ratio growth in agreement with expectations. A
significant fraction of the \emph{even} isotopes of Ba found in
old Galactic stars (the thick disk stars), $\sim67$\%, is in
contrast to the prediction of the "classical" model of the
$s-$process and favors the value predicted by the "stellar" models
of Arlandini et al. (1999) and Travaglio et al. (1999).
 \keywords{Line: formation -- Nuclear reactions, nucleosynthesis, 
abundances -- Stars: abundances -- Stars: late-type -- Galaxy: evolution}
}
 \maketitle

\section{Introduction}

The elements heavier than the iron peak are mainly produced
through neutron capture reactions in two main processes, the
$s-$process (slow) and $r-$process (rapid). As supported by many
observational and theoretical results (Travaglio et al.
\cite{eu99}, and references therein), $s-$nuclei are mainly
synthesized during the thermally pulsing asymptotic giant branch
phase of low-mass stars (2 - 4 M$_\odot$). The $r-$process is
associated with explosive conditions in SNeII (see Thielemann et
al. \cite{r2002} for a general review). Following the pioneering
observational study of Spite \& Spite (\cite{Spite}), Truran
(\cite{Truran}) has put forward the suggestion that most (if not
all) neutron capture elements in very metal-poor stars are of 
$r-$process origin. Much observational effort was invested in
testing this idea (for review, see Truran et al. \cite{19}). 
To understand the relative importance
of the two neutron capture synthesis mechanisms throughout the
Galaxy history, stellar element abundance comparisons are made
between elements whose solar system isotopic abundances are
dominated by the $s-$process and those due mainly to the
$r-$process. The europium-to-barium abundance ratio is
particularly sensitive to whether nucleosynthesis of the heavy
elements occurred by the $s-$ or $r-$process. Other important
indicators are the [Eu/La] and [Nd/Ba] abundance ratios. The most
recent determinations of the [Eu/Ba] abundance ratios in the
samples of metal-poor stars were made by McWilliam (\cite{mcw98}),
Burris et al. (\cite{Burris}), Fulbright (\cite{Fulb00}),
Mashonkina \& Gehren (\cite{euba}, \cite{mash03}), Mishenina \&
Kovtyukh (\cite{mish_eu}) and Bensby et al. (\cite{bensby05}). The
[Eu/La] abundance ratios in 159 giant and dwarf stars have been
measured by Simmerer et al. (\cite{la2004}); the [Nd/Ba] abundance
ratios in 46 stars have been derived by Mashonkina et al.
(\cite{mash_nd}). Stellar element abundance ratios have to be
compared with theoretical predictions. However, at present, the
nuclesynthesis theory is not able to predict the yields of the
$r-$process. Based on the hypothesis that the $r-$contributions in
the solar system are of primary origin, the $r-$process abundance
fractions are obtained by subtracting the $s-$contributions at the
epoch of solar formation, estimated with the use of reliable
$s-$model, from the solar abundances. One refers to them as
$r-$residuals. Two approaches are used to calculate  the
$s-$process abundance fractions. The "classical" approach relies
upon reproducing the product of $n-$capture cross-section and
$s-$process abundance (the "$\sigma$N" curve). "Stellar"
$s-$process models are based on nucleosynthesis computations in
low- and intermediate-mass AGB stars. 
  The recent ``classical'' (Arlandini et al. \cite{rs99}) and ``stellar'' 
(Arlandini et al. \cite{rs99}; Travaglio et al. \cite{eu99}) models give 
consistent results within the respective uncertainties for the $s-$process 
contributions to the most abundant Ba and Eu isotopes in the solar system 
matter, with the only exception being \iso{138}{Ba}. Its 
$r-$residual equals 14\%\, to 16\%\, in the ``stellar'' models and 0\%\, 
 in the ``classical'' model.
Since \iso{138}{Ba}\, is the most abundant solar Ba isotope, the $r-$residual 
of \emph{total} Ba is larger in the ``stellar'' model by a factor of 2.4 
than in the ``classical'' one. As a result, the ``stellar'' and ``classical'' models lead to significantly different values of  
the solar abundance ratio of Eu to Ba contributed by the $r-$process: taken 
relative to the total abundances, [Eu/Ba]$_r$ = 0.69 and 
[Eu/Ba]$_r$ = 1.06, respectively. The ``stellar'' model predicts the isotope 
abundance ratio \iso{135}{Ba} : \iso{137}{Ba} : \iso{138}{Ba}
= 26 : 20 : 54 in a pure $r-$process nucleosynthesis, while no Ba even isotope
is predicted in the ``classical'' model.

To reconstruct the evolutionary history of neutron-rich
elements in the Galaxy it is, thus, very important to disentangle
the fractions of Ba \emph{even and odd isotopes} and inspect their
abundance as a function of Galactic age. Isotopic shifts of the
barium lines are negligible compared to the line width. A
determination of the Ba even-to-odd isotope abundance ratio in
stars becomes possible due to the significant hyperfine structure
(HFS) affecting the \ion{Ba}{ii}\, resonance lines of the odd
isotopes. 
In total, the resonance line $\lambda4554$ has 15
components spread over 58 m\AA. The larger the fraction of the odd isotopes, 
the stronger the HFS broadening of $\lambda4554$ and the larger the energy  
absorbed in this line. The total width of the patterns of
the \ion{Ba}{ii}\, subordinate lines, $\lambda5853$, $\lambda6141$,
and $\lambda6496$, is much smaller, 8 m\AA, 8 m\AA, and 23 \AA,
correspondingly. 

Cowley \& Frey (\cite{cowley}) pointed out the importance of
accounting for HFS in Ba abundance determinations; however, the
first attempt to estimate the fraction of the odd isotopes of Ba
in four metal-poor stars was made by Magain \& Zhao
(\cite{Mag93a}). They used the \ion{Ba}{ii}\, subordinate lines,
$\lambda5853$ and $\lambda6496$, to derive the total Ba abundance.
The Ba abundance was then deduced from the $\lambda4554$ line for
various isotopic mixtures, the proportion of the odd isotopes
being changed until agreement with the subordinate lines was
obtained. Magain \& Zhao found an enhancement of the odd isotopes
in these stars as compared to solar system matter, in agreement
with the expectations. Using the same approach, Mashonkina et al.
(\cite{Mash99}) did not find significant deviation of the Ba
isotopic mixture in the halo star G246-38 ([Fe/H] = --2.20) from
the solar one. In their recent study, Mashonkina et al.
(\cite{mash03}) show a distinction between the different Galactic
stellar populations with respect to the Ba even-to-odd isotope ratio. The
halo stars reveal, on average, equal amounts of the odd and even
isotopes of Ba; a mean ratio 65 : 35 ($\pm$10\%) was obtained for
the sample of thick disk stars, whereas the solar ratio 82 : 18
(Anders \& Grevesse \cite{AG}) adjusts to observations of the
\ion{Ba}{ii}\, lines in the sample of thin disk stars.

Magain \& Zhao (\cite{Mag93}) and Magain (\cite{Mag95}) have
suggested a different method based on measuring the broadening of
the \ion{Ba}{ii}\, $\lambda$ 4554 line. They have applied it to
the extremely metal-poor star HD\,140283 and found the low
fraction of the odd isotopes, 8\%, that is consistent with a pure
$s-$process production of barium. Using the same method and new
high resolution (R $\simeq$ 200000), high signal-to-noise ratio
(S/N $\simeq$ 550) spectra of HD\,140283, Lambert \& Allende Prieto
(\cite{iso140283}) draw the opposite conclusion. They find that
the $r-$process mixture of the Ba isotopes predicted by the
"stellar" model of Arlandini et al. (\cite{rs99}) provides a fair
fit to the observed \ion{Ba}{ii}\, $\lambda$ 4554 profile.

 This study is intended to inspect a history of Ba isotopic fractions in 
the Galaxy and, thus, a relative importance of the $r-$ and $s-$process in 
heavy element production through the Galaxy evolution. 
We determine the abundance ratios of even to odd isotopes of Ba in 
the 25 selected stars representing the older (thick disk) and 
younger (thin disk) Galactic stellar populations. We use the
method applied in our previous papers. The results are based on
the high resolution ($\sim 60\,000$) high signal-to-noise ratio
(S/N $\ge 200$) spectra and non-local thermodynamical equilibrium
(NLTE) line formation.

The paper is organized as follows. In Sect.~\ref{method} we
discuss an accuracy of the used atomic parameters, oscillator
strengths and van der Waals damping constants $C_6$, and check the
investigated \ion{Ba}{ii}\, lines with respect to blending lines.
We test our method on solar \ion{Ba}{ii}\, lines. Stellar sample,
observations, and stellar parameters are described in
Sect.~\ref{obs}. In Sect.~\ref{isotope} we present the obtained Ba 
even-to-odd isotope ratio in stars and investigate their errors caused by 
uncertainties of atomic data and stellar parameters. In Sect.~\ref{concl}, 
stellar fractions of the odd isotopes of Ba are inspected as indicators of the 
$r/s-$process nucleosynthesis, and conclusions are given. At the end, 
the outlook for a determination of Ba isotopic fractions in very 
metal-poor stars is estimated.

\section{Method of calculations}\label{method}

Homogeneous blanketed model atmospheres computed with the MAFAGS
code (Fuhrmann et al. \cite{Fuhr1}) are used in the analysis of both
solar and stellar spectra.

\subsection{Atomic parameters}\label{data}

In the metallicity range of our stellar sample, not only the
resonance line, but also one or both of the used subordinate lines
either lie on the damping part of the curve of growth or are
saturated. In the first turn, our analysis requires very accurate
atomic parameters.

As in our previous analysis, in this study we use the absolute 
\emph{oscillator strengths} from  Reader et al.  
(\cite{WM}): $\log gf (\lambda\,4554) = 0.162$, $\log gf
(\lambda\,5853) = -1.0$, and $\log gf (\lambda\,6496) = -0.377$.
The most recent measurements of Davidson et al. (\cite{fba92}) give 
$\log gf (\lambda\,4554) = 0.140$, $\log gf
(\lambda\,5853) = -0.91$, and $\log gf (\lambda\,6496) = -0.407$.  
We show below that 
applying the data of Davidson et al., we cannot agree when fitting parameters 
of the solar \ion{Ba}{ii}\, $\lambda\,6496$ and 
$\lambda\,5853$ lines.
We do not use the third line of the \ion{Ba}{ii} $5d - 6p$ multiplet,
$\lambda\,6141$ which is blended with the strong Fe\,I line
(Fig.~\ref{solar}).

In our previous studies, we used the \emph{data on hyperfine
structure} components of the \ion{Ba}{ii}\, $\lambda\,4554$
calculated by Biehl (\cite{hfs15}) on the base of Brix \&
Kopfermann (\cite{hfs}) measurements. Having reviewed the
literature for more recent data, we found that the new fine
structure constants of \iso{135}{Ba}\, and \iso{137}{Ba}\,
obtained by Blatt \& Werth (\cite{hfs137}) and Becker \& Werth
(\cite{hfs135}) lead to a negligible change of the wavelength
separations by no more than 1\%\, for the HFS components of the
resonance line. In order to have common input data with other
authors that makes a direct comparison of the results possible, in
this study all calculations of $\lambda\,4554$ are performed using
the HFS pattern published by McWilliam (\cite{mcw98}). We
emphasize that the line profiles computed with the data of Biehl
and McWilliam become identical after their relative shift by 
0.001~\AA. Table~\ref{hfs4554} contains the list of the HFS components
of the \ion{Ba}{ii} $\lambda\,4554$. The product of $f_{ij}$ and 
fractional isotope abundance $\epsilon$ is given for the Ba isotope mixture in
the solar system matter, \iso{134}{Ba} :
\iso{135}{Ba} : \iso{136}{Ba} : \iso{137}{Ba} : \iso{138}{Ba} =
2.4 : 6.6 : 7.9 : 11.2 : 71.8 (Anders \& Grevesse \cite{AG}) and
that for a pure $r-$process production of Ba, \iso{135}{Ba} :
\iso{137}{Ba} : \iso{138}{Ba} = 26 : 20 : 54 (Arlandini et al.
\cite{rs99}).  Total $f_{ij}$ = 0.727. 
The 23~m\AA\, wide patterns of the $\lambda6496$
line are considered as a four component model according to Rutten
(\cite{Rut}). Variation in fractional isotope abundances between
the solar and pure $r-$process cases has negligible effect on
the $\lambda6496$ line profile, and we use $f\times\epsilon$ 
computed for the solar mixture of Ba
isotopes. We neglect the hyperfine structure of $\lambda5853$.

\begin{table}
\caption{Atomic data for the HFS components of the \ion{Ba}{ii}
$\lambda\,4554$. $I$ is the relative strength,  
the product of oscillator strength and fractional isotope abundance,
 $f\times\epsilon$, is given for two Ba isotope mixtures 
corresponding to  
the solar system matter (column ``solar'') and $r-$process yields 
predicted by Arlandini et al. (1999)}
\label{hfs4554}
\begin{center}
\begin{tabular}{ccll}
\hline
 ~~$\lambda,\AA$ &   $I$   & \multicolumn{2}{c}{$f\times\epsilon$} \\
\cline{3-4}
  &      & solar & $r-$process \\
\hline
\iso{134}{Ba} & & & \\
4554.032 & 1.000 & 0.0176  & ~~~- \\
\iso{135}{Ba} & & & \\
4554.003 & 0.1562 & 0.0075 &  0.0292 \\
4554.004 & 0.1562 & 0.0075 &  0.0292 \\
4554.005 & 0.0625 & 0.0030 &  0.0117 \\
4554.051 & 0.4375 & 0.0209 &  0.0818 \\
4554.054 & 0.1562 & 0.0075 &  0.0292 \\
4554.055 & 0.0313 & 0.0015 &  0.0058 \\
\iso{136}{Ba} & & & \\
4554.033 & 1.000 & 0.0568  & ~~~- \\
\iso{137}{Ba} & & & \\
4553.999 & 0.1562 & 0.0129 &  0.0232 \\
4554.001 & 0.1562 & 0.0129 &  0.0232 \\
4554.002 & 0.0625 & 0.0052 &  0.0093 \\
4554.054 & 0.4375 & 0.0360 &  0.0650 \\
4554.056 & 0.1562 & 0.0129 &  0.0232 \\
4554.057 & 0.0313 & 0.0025 &  0.0046 \\
\iso{138}{Ba} & & & \\
4554.034 & 1.000 & 0.521 &  0.392 \\
\hline
\end{tabular}
\end{center}
\end{table}

The investigated \ion{Ba}{ii}\, lines are strongly affected by
\emph{van der Waals damping}. We apply the van der Waals damping 
constant $\log C_6$ = --31.65 to $\lambda\,4554$. If $\log C_6$ is fixed, 
solar Ba abundance is immediately determined from the fitting of the solar 
$\lambda\,4554$ line wings.
Using the MAFAGS solar model atmosphere, we find  
$\eps{\odot Ba} = 2.21$ \footnote{We refer to abundances 
on the usual scale where $\eps{H} = 12$.}. 
This value agrees within error bars with the 
meteoritic Ba abundance given by Anders \& Grevesse (\cite{AG}, 
$\eps{met,Ba} = 2.21\pm0.03$) and Grevesse et al. (\cite{met96}, 
$\eps{met,Ba} = 2.22\pm0.02$). Recent determinations by Asplund et al.
(\cite{met05}) lead to the smaller meteoritic ($\eps{met,Ba} = 2.16\pm0.03$) 
and solar ($\eps{\odot Ba} = 2.17\pm0.07$) Ba abundance. For comments, see Note added in proofs.

The Van der Waals damping constant $\log C_6$ of the
subordinate lines was obtained empirically from the fitting of the
solar line profiles under the requirement that unique set of
fitting parameters (Ba abundance, microturbulence value,
and $C_6$ value) must adjust observations of the spectral lines
belonging to the same multiplet, $5d - 6p$. Far wings of
$\lambda\,6496$ impose the upper limit of $\log C_6$,  provided that 
solar Ba abundance is given. We adopt $\eps{\odot Ba} = 2.21$. In this 
case, $\log C_6$ of $\lambda\,6496$ cannot be larger than --31.3. Applying
this value to $\lambda\,5853$ and based on NLTE analysis we obtain
a microturbulence value $\Vmic$ = 0.85~\kms. The best fit of
each line of the multiplet $5d - 6p$ is achieved at $\eps{Ba}$ =
2.21, $\log C_6$ = --31.3, and $\Vmic$ = 0.85~\kms\,
(Fig.~\ref{solar}). 
In Sect.~\ref{isotope} we will discuss an effect of 
variation in $\log C_6$ on Ba isotope ratio in stars. Analysis of the solar 
$\lambda\,6496$ and $\lambda\,5853$ line profiles was also performed with the 
oscillator strengths from Davidson et al. (\cite{fba92}). Solar Ba abundance  
was fixed at $\eps{\odot Ba} = 2.21$. 
Fitting parameters of two lines turn out to be different: $\lambda\,6496$ 
 gives $\log C_6$ = --31.2 and $\Vmic$ = 0.9~\kms, while the significantly 
lower microturbulence value $\Vmic$ = 0.7~\kms\, is required to fit the 
$\lambda\,5853$ line profile using $\log C_6$ = --31.2. 

The \ion{Ba}{ii} $\lambda\,4554.034\,\AA$ line is blended by the
Cr~I $\lambda\,4553.945$ and \ion{Zr}{ii} $\lambda\,4553.934$
lines in the blue line wing and by the CH molecular line at
4554.142~\AA\, in the red line wing (Fig.~\ref{sun_4554iso}). 
 We have requested the Vienna Atomic Line 
Data base (VALD, Kupka et al. \cite{vald}) and the NIST atomic
spectra database (http://physics.nist.gov/PhysRefData) to search for 
atomic parameters of blending atomic lines. The
same value $\log gf = -0.73$ based on measurements of Wujec \& 
Weniger (\cite{cr1fij}) is given in NIST and VALD for the Cr~I 
$\lambda\,4553.945$. NIST does not contain the \ion{Zr}{ii} 
$\lambda\,4553.934$ line, but VALD provides 
 $\log gf = -0.57$, taken from Cowley \& Corliss 
(\cite{zr2fij}), for this line.
The oscillator strength of the CH $\lambda\,4554.142$ molecular line
was fitted to reproduce the observed blend profile; we have
obtained $\log gf = -3.2$. The \ion{Ba}{ii}
$\lambda\,5853.675\,\AA$ line is blended by the Fe~I
$\lambda\,5853.682$ line (Fig.~\ref{solar}).  Based on calculations 
of Kurucz \& Bell (\cite{kur95}), VALD gives $\log gf$
(Fe~I $\lambda\,5853.682$) = --2.148. Recently, Kurucz 
(http://cfaku5.cfa.harvard.edu) calculated the new value $\log gf$ = 
--2.371 for this line. We note that the predicted oscillator strength 
of Fe~I $\lambda\,5853.682$ has been reduced by a factor of 2.3, compared to 
the first estimate of Kurucz \& Peytremann (\cite{kur75}), $\log gf$ = 
--2.0. This line
is absent in the NIST database. The \ion{Ba}{ii}\, 
$\lambda\,6141.713\,\AA$ is strongly blended by the Fe~I
$\lambda\,6141.732$ line and not used in further analysis of
stellar spectra. With $\log gf$ (Fe~I $\lambda\,6141$) =
--1.61 taken from the NIST database, we can achieve a good fitting of
the solar blend profile (Fig.~\ref{solar}). VALD gives the larger 
oscillator strength, $\log gf$ = --1.459, for 
Fe~I $\lambda\,6141$. The contribution of blending lines is treated using
the LTE assumption. The chemical abundances are taken from Anders
\& Grevesse (\cite{AG}) in the solar case, and we assume that in
the investigated stars, Cr abundance follows the iron one with [Cr/Fe]
= 0, and Zr abundance follows the barium one with [Zr/Ba] = 0.

\subsection{NLTE line formation}

Our results are based on NLTE line formation for \ion{Ba}{ii}. The
method of NLTE calculations was developed and described earlier
(Mashonkina \& Bikmaev \cite{Mash96}; Mashonkina et al.
\cite{Mash99}). We use a revised version of the DETAIL program
(Butler \& Giddings \cite{detail}) based on the accelerated lambda
iteration method to solve the coupled radiative transfer
and statistical equilibrium equations and the SIU code
(www.usm.uni-muenchen.de/people/reetz/siu.html) to compute the
synthetic line profiles.

\begin{figure}
\resizebox{88mm}{!}{\includegraphics{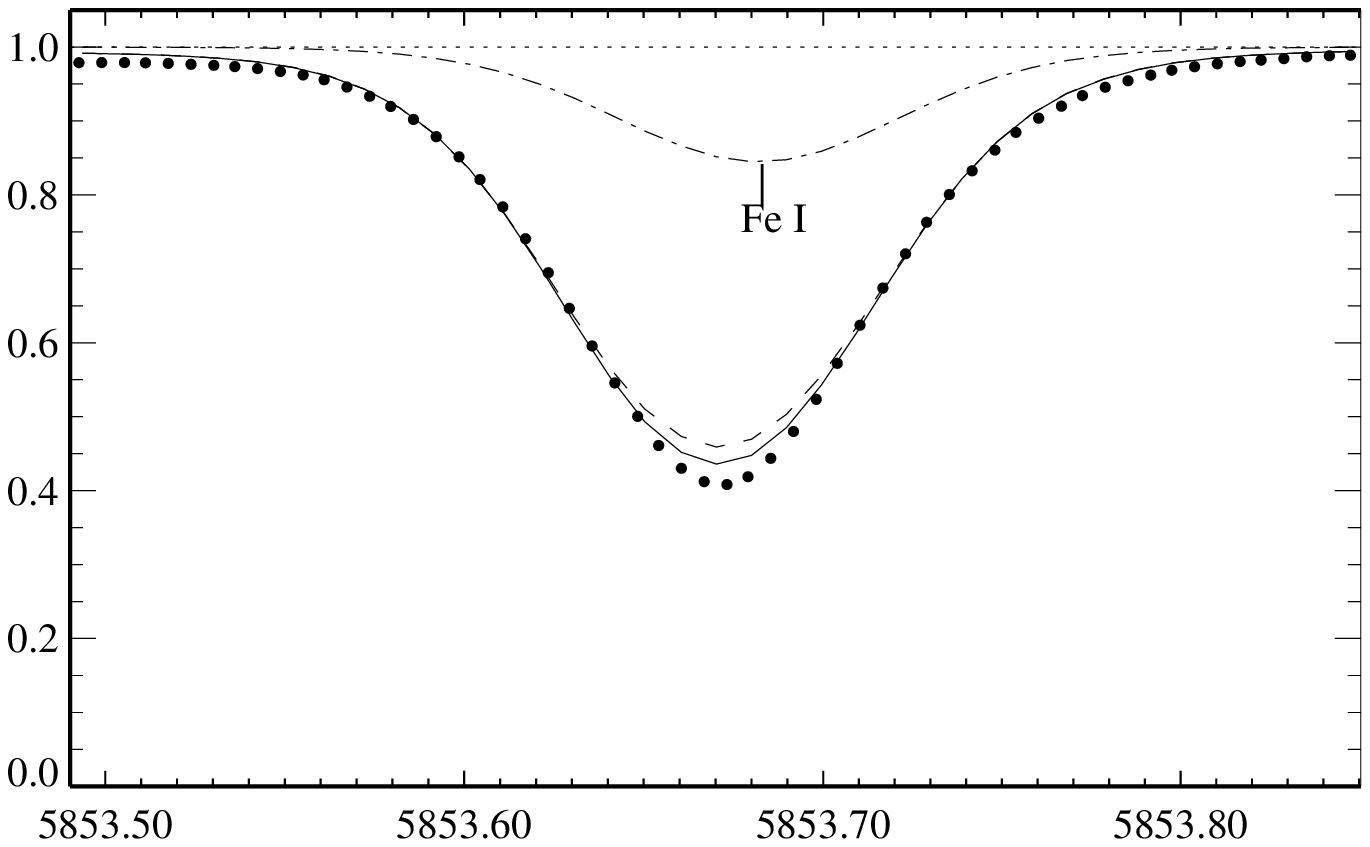}} \vspace{-0mm}
\resizebox{88mm}{!}{\includegraphics{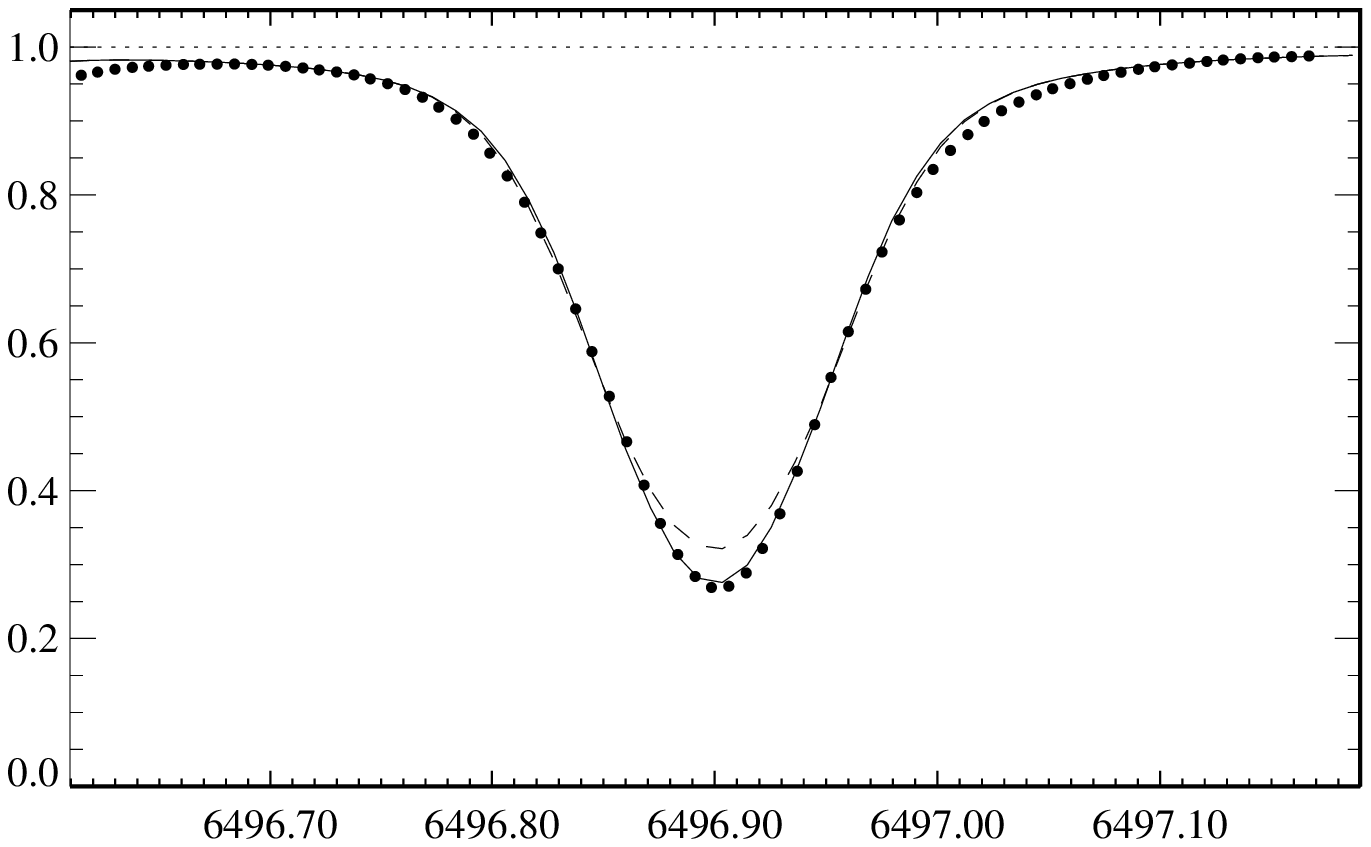}} \vspace{-0mm}
\resizebox{88mm}{!}{\includegraphics{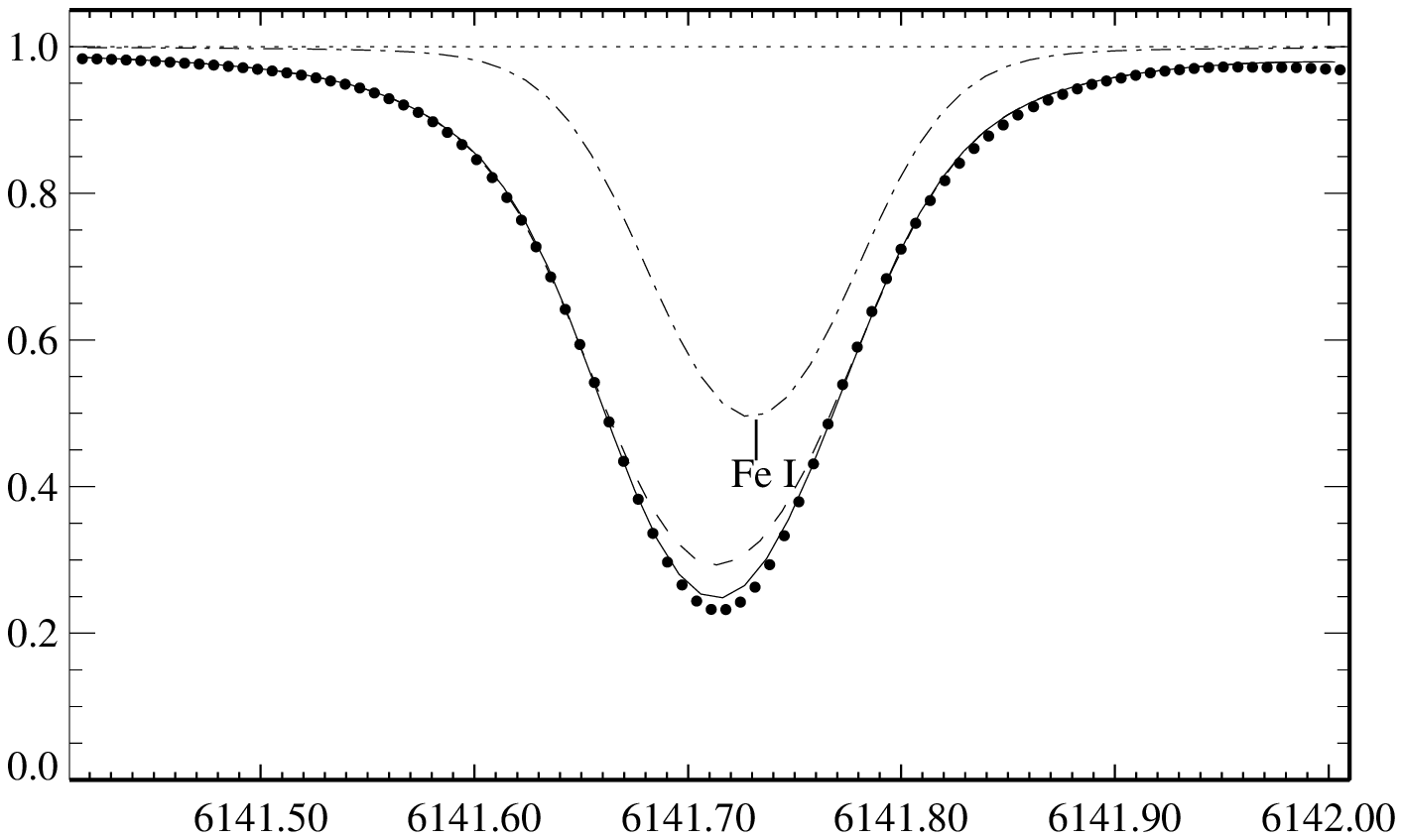}} \vspace{-3mm}
\caption[]{The best fits of the solar flux Ba~II subordinate line
profiles (Kurucz et al. 1984, bold dots) achieved using the NLTE
approach (continuous line) and LTE assumption (dashed line). Solar
Ba abundance $\eps{Ba}$ = 2.21; $\log C_6 = -31.3$. NLTE profiles:
each is calculated using $\Vmic$ = 0.85~\kms. In the LTE case,
$\Vmic$ = 0.95~\kms\, ($\lambda\,5853$), 1.05~\kms\,
 ($\lambda\,6496$), and 1~\kms\, ($\lambda\,6141$). Blending lines are shown in
each panel by a dash-dotted line. See text for more details.}
\label{solar}
\end{figure}

In Fig.~\ref{solar}, we present the best fits of the observed
\ion{Ba}{ii}\, subordinate line profiles in the solar flux atlas
of Kurucz et al. (\cite{Atlas}) achieved using the NLTE
approach and LTE assumption. We emphasize that the unique value
of Ba abundance, $\eps{Ba}$ = 2.21, adjusts observations of all
these lines. The same value, $\log C_6 = -31.3$, is used for each
line in both the NLTE and LTE calculations. The same microturbulence value,
$\Vmic$ = 0.85~\kms, is required when the NLTE approach is applied
while different values in the LTE case are used: $\Vmic$ = 0.95~\kms\, 
for $\lambda\,5853$, $\Vmic$ = 1.05~\kms\, for $\lambda\,6496$, and 
$\Vmic$ = 1~\kms\, for $\lambda\,6141$. Our synthetic flux profiles 
are convolved with a
profile that combines a rotational broadening of 1.8 \kms\, and
broadening by macroturbulence with a radial-tangential profile.
The macroturbulence values $\Vmac$ = 3.1~\kms\, and 3.0~\kms\, are
required for $\lambda\,5853$ in the NLTE and LTE cases,
respectively. The corresponding values for $\lambda\,6496$ are
2.3~\kms\, and 2.1~\kms\, and for $\lambda\,6141$, in both cases, $\Vmac$ =
2.9~\kms.

\subsection{Can we derive the solar Ba isotope abundance ratio?}\label{4554}

\begin{figure}
\resizebox{88mm}{!}{\includegraphics{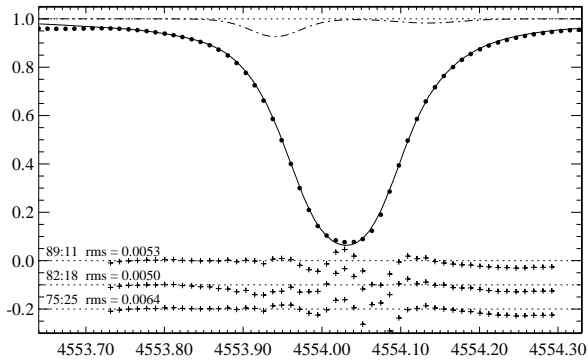}}
\vspace{-3mm} \caption[]{Synthetic NLTE flux profile of the Ba~II
resonance line corresponding to the Ba even-to-odd isotope ratio of
82 : 18 (continuous line)
compared with the solar flux profile (bold dots, Kurucz et al. 1984).
 The differences between observed
and calculated spectra, (O - C), multiplied by a factor of 5, are
shown for various even-to-odd isotope ratios in the lower part of
figure. Blending lines are shown by a dash-dotted line. See text for
more details} \label{sun_4554iso}
\end{figure}

Assuming the \emph{solar system matter} Ba isotopic fractions 
\iso{134}{Ba} : \iso{135}{Ba} : \iso{136}{Ba} : \iso{137}{Ba} :
\iso{138}{Ba} = 2.4 : 6.6 : 7.9 : 11.2 : 71.8, total Ba abundance
$\eps{Ba}$ = 2.21, $\log C_6 = -31.65$, and using $\Vmic$ = 1~\kms, 
we can achieve a good fitting
of the solar $\lambda\,4554$ profile. In Fig.~\ref{sun_4554iso}, we
show the best fit obtained by applying $\Vmac$ = 3.5~\kms. 

The value $\Vmic$ = 1~\kms\, obtained from the resonance line does
not agree with the value $\Vmic$ = 0.85~\kms\, found from the
subordinate lines. We are aware that a phenomenological
description of the velocity field using a microturbulence value is
a rough approximation and that it fails to treat spectral lines formed
over a wide range of depths. The $\lambda\,4554$ line core is formed
in very outer layers near $\log \tau_{5000} = -4.7$, while
radiation of the subordinate lines comes from the deeper layers,
below $\log \tau_{5000} = -2.8$ for $\lambda\,6496$ and below
$\log \tau_{5000} = -1.9$ for $\lambda\,5853$. We have tried to
answer the question of whether we are able to derive a \emph{solar
atmosphere} Ba isotope abundance ratio from the HFS affecting the
$\lambda\,4554$ line profile without a reference to the
subordinate lines. In addition to the even-to-odd isotope ratio 82 : 18 
corresponding to solar system matter, a 
fitting of the solar $\lambda\,4554$ line
profile was made assuming the ratios 89 : 11 and 75 : 25. 
We suppose that
the proportion \iso{135}{Ba} : \iso{137}{Ba} does not change. It
is not the case if the $r/s-$process contribution to Ba varies.
However, as can be
seen from Table~\ref{hfs4554}, the $\lambda\,4554$ line profile
depends on the total fractional abundance of the odd
isotopes rather than on the proportion \iso{135}{Ba} : \iso{137}{Ba}. Our
test computations for the isotope mixture \iso{135}{Ba} :
\iso{137}{Ba} : (\iso{134}{Ba} + \iso{136}{Ba} + \iso{138}{Ba}) = 2
: 9 : 89 predicted by Arlandini et al. (\cite{rs99}) for a pure
$s-$process production of Ba give the $\lambda\,4554$ line profile
fully consistent with that for the ratio \iso{135}{Ba} :
\iso{137}{Ba} : (\iso{134}{Ba} + \iso{136}{Ba} + \iso{138}{Ba}) =
5.5 : 5.5 : 89. In calculations with various even-to-odd isotope
ratio, a microturbulence value was a free parameter, and in each
case the best fit was found from the minimum root mean square
($rms$) difference between observed and calculated spectra. 
$\Vmac$ was fixed at 3.5~\kms. The minimum $rms$ is achieved
at $\Vmic$ = 1.15~\kms\, for the even-to-odd isotope ratio 89 : 11,
and $\Vmic$ = 0.9~\kms\, for 75 : 25. In the latter case, a
shift of the observed profile by --0.0008~\AA\, is required.
Fig.~\ref{sun_4554iso} shows the obtained 
differences between observed and calculated spectra, (O - C), multiplied 
by a factor of 5, for various isotope ratios. 

The $rms$ values quoted in Fig.~\ref{sun_4554iso} are very similar 
for various isotope mixtures. The obtained $\Vmic$
values 0.9~\kms\, to 1.15~\kms\, are within the interval of
widely used solar microturbulence values, and no value can be
preferred. The (O - C) values are the largest in the line core
within $\pm0.07~\AA$ from the line center, independent of isotope
ratio. The $\lambda\,4554$ line core is most probably influenced
by a non-thermal and depth-dependent chromospheric velocity field
that is not part of our solar model. We note that the
chromospheric temperature rise, as well as the horizontal temperature
inhomogeneity seen as solar granulation, is expected to be less
important because \ion{Ba}{ii}\, is the dominant ionization stage.

We conclude from these results that a precise determination of the
proportion of the even and odd isotopes of Ba in the solar
atmosphere matter requires a correct treatment of the atmospheric
velocity field that is only possible on the basis of hydrodynamic
calculations. In metal-poor stars, the $\lambda\,4554$ line 
forms in the deeper atmospheric layers compared to the Sun and in 
smaller extension is expected to be affected by hydrodynamic
phenomena of the very surface layers.

\section{Stellar sample, observations, and stellar
parameters}\label{obs}

\begin{figure}
\resizebox{88mm}{!}{\includegraphics{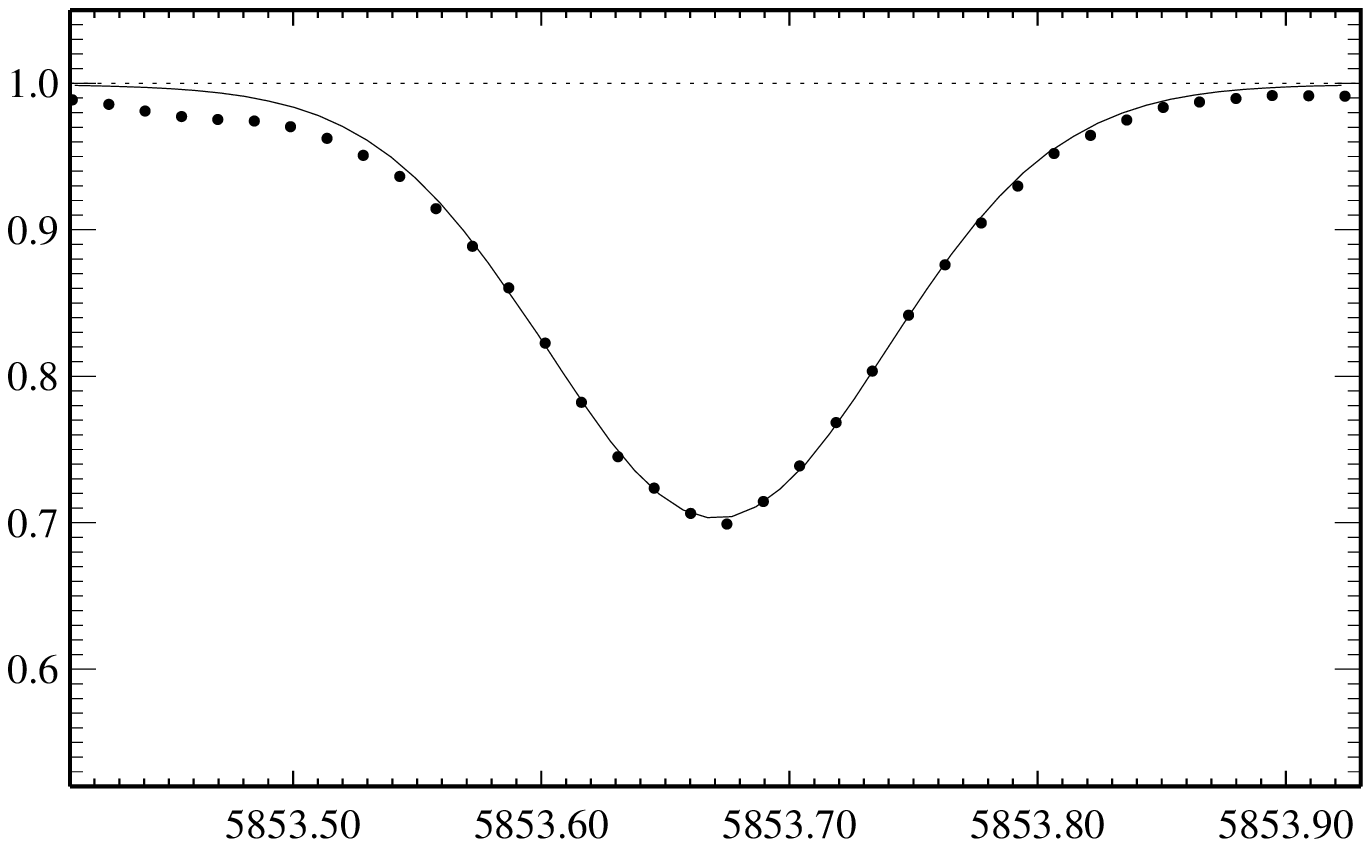}} \vspace{-0mm}
\resizebox{88mm}{!}{\includegraphics{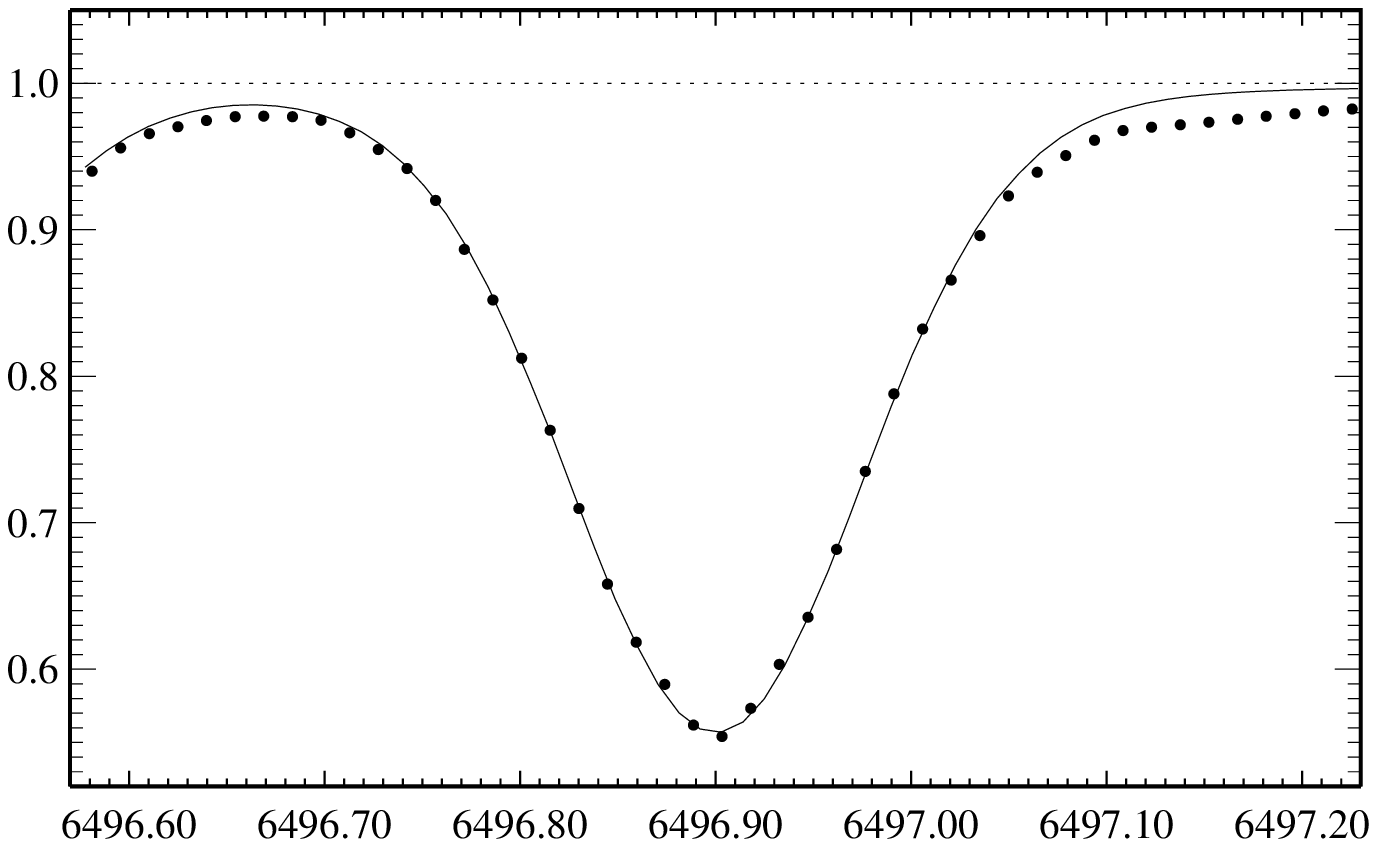}} \vspace{-0mm}
\caption[]{Synthetic NLTE (continuous line) flux profiles of the
Ba~II subordinate lines compared with the observed FOCES spectra
(bold dots) of HD\,10519. The same value [Ba/H] = --0.69 is
obtained from both lines} \label{star_sub}
\end{figure}

All the 25 stars were selected from Fuhrmann's (\cite{Fuhr04})
list. Each star was observed with a resolution of $\sim 60\,000$,
at least twice, and for each star both the \ion{Ba}{ii}\,
subordinate lines, $\lambda$5853 and $\lambda$6496, are available
in its spectra. Spectra were obtained by Klaus Fuhrmann using the
fiber optics Cassegrain echelle spectrograph FOCES at the 2.2m
telescope of the Calar Alto Observatory in 1997 - 2000. The
signal-to-noise ratio is $200$ or higher in the spectral range
$\lambda > 4500$ \AA. Typical line profiles are seen in 
Figs.~\ref{star_sub} and \ref{star_4554}.

The list of stars under investigation is given in
Table~\ref{startab}. Stellar parameters have been determined
spectroscopically by Fuhrmann (\cite{Fuhr04}). Their errors are
estimated as $\Delta\Teff = 80$\,K, $\Delta\log g = 0.1$,
$\Delta$[Fe/H] = 0.1~dex, and $\Delta\Vmic = 0.2$~\kms. We use also
his identification of a membership of individual stars in the
particular stellar population that is based on the star's
kinematics, metallicity, $\alpha-$enhancement, and age. Our sample
includes 15 thick disk stars, 9 thin disk stars, and one halo star.

\begin{table}[htbp]
\caption{ Stellar parameters, Ba abundances, and the fractions of
the odd isotopes of Ba (column {\it odd}, in \%) of the selected
sample. The column {\it low - up} contains the lower and upper
limits of that fraction.  In the column ``Note'', the notations 0, 1, 
and 2 refer to the thin disk, thick disk, and halo stars, respectively. 
$\Vmic$ is given in \kms} \label{startab} \tabcolsep1.2mm
\begin{center}
\begin{tabular}{rcccrrccc}
 & & & & & & & & \\
\hline \multicolumn{1}{c}{HD} & $\Teff$ & $\log g$ & $\Vmic$ &
[Fe/H] & [Ba/H] &{\it odd} &  {\it low - up} & Note \\
\hline
    3795   & 5370  & 3.82 & 1.0 & --0.64 & --0.58 & 39 & 34 - 44 & 1 \\
    4614   & 5940  & 4.33 & 1.0 & --0.30 & --0.23 & 24 & 17 - 31 & 0 \\
    9407   & 5660  & 4.42 & 0.9 &   0.03 & --0.01 & 13 &  7 - 18 & 0  \\
    10519  & 5710  & 4.00 & 1.1 & --0.64 & --0.69 & 42 & 38 - 46 & 1  \\
    10697  & 5610  & 3.96 & 1.0 &   0.10 &   0.14 & 11 &  4 - 17 & 0  \\
    18757  & 5710  & 4.34 & 1.0 & --0.28 & --0.38 & 31 & 26 - 36 & 1  \\
    22879  & 5870  & 4.27 & 1.2 & --0.86 & --0.88 & 31 & 28 - 34 & 1  \\
    30649  & 5820  & 4.28 & 1.2 & --0.47 & --0.57 & 30 & 27 - 33 & 1  \\
    37124  & 5610  & 4.44 & 0.9 & --0.44 & --0.55 & 29 & 25 - 34 & 1  \\
    52711  & 5890  & 4.31 & 1.0 & --0.16 & --0.09 & 16 & 11 - 22 & 0  \\
    55575  & 5890  & 4.25 & 1.0 & --0.36 & --0.39 & 23 & 20 - 26 & 0  \\
    62301  & 5940  & 4.18 & 1.2 & --0.69 & --0.75 & 32 & 28 - 36 & 1  \\
    64606  & 5320  & 4.54 & 1.0 & --0.89 & --1.00 & 35 & 30 - 40 & 1  \\
    65583  & 5320  & 4.55 & 0.8 & --0.73 & --0.77 & 39 & 34 - 44 & 1  \\
    68017  & 5630  & 4.45 & 0.9 & --0.40 & --0.50 & 33 & 25 - 38 & 1  \\
    69611  & 5820  & 4.18 & 1.2 & --0.60 & --0.72 & 28 & 24 - 32 & 1  \\
    102158 & 5760  & 4.24 & 1.1 & --0.46 & --0.59 & 35 & 29 - 43 & 1  \\
    103095 & 5110  & 4.66 & 0.8 & --1.35 & --1.41 & 42 & 36 - 46 & 2  \\
    112758 & 5240  & 4.62 & 0.7 & --0.43 & --0.53 & 27 & 16 - 37 & 1  \\
    114762 & 5930  & 4.11 & 1.2 & --0.71 & --0.83 & 37 & 31 - 42 & 1  \\
    117176 & 5480  & 3.83 & 1.0 & --0.11 & --0.09 & 25 & 11 - 38 & 0  \\
    121560 & 6140  & 4.27 & 1.2 & --0.43 & --0.35 & 27 & 22 - 32 & 0  \\
    132142 & 5240  & 4.58 & 0.7 & --0.39 & --0.46 & 33 & 29 - 37 & 1  \\
    134987 & 5740  & 4.25 & 1.0 &   0.25 &   0.17 & 18 & 13 - 21 & 0  \\
    168009 & 5785  & 4.23 & 1.0 & --0.03 & --0.07 & 16 & 12 - 21 & 0  \\
\hline
\end{tabular}
\end{center}
\end{table}  %

\section{Results} \label{isotope}

\subsection{Stellar total Ba abundances}

Ba abundances were derived from the \ion{Ba}{ii} $\lambda\,5853$
and $\lambda\,6496$ line profile fitting, and then the average
value was calculated. The [Ba/H] values calculated using solar Ba
abundance $\eps{Ba}$ = 2.21 are presented in Table~\ref{startab}.
We note that in this paper we revise our earlier determinations
(Mashonkina \& Gehren \cite{eubasr}) of Ba abundance in the
investigated stars. Contrary to the previous analysis, we use
the same value $\log C_6$ for both subordinate lines and treat the
$\lambda\,5853$ blend taking into account both the \ion{Ba}{ii}
and Fe~I lines (see Sect. \ref{data}). A reduction of the van der
Waals damping constant from $\log C_6 = -30.6$ for
$\lambda\,5853$ and $\log C_6 = -31.2$ for $\lambda\,6496$ used 
in 2001 to our
present estimate $\log C_6 = -31.3$ (Sect. \ref{data}) leads to an
increase of derived Ba abundance, while accounting for absorption
by the blending of the Fe~I line leads to a decrease of Ba abundance from the
$\lambda\,5853$ blend. As a result, in the two most metal-poor
stars of our sample (HD\,103095 and HD\,64606), the revised Ba
abundance is smaller (van der Waals broadening of the subordinate
lines is weak and blending of \ion{Ba}{ii} $\lambda\,5853$
prevails) by 0.01~dex compared to our previous data. Also in the
remaining stars, the revised Ba abundance has become larger by 0.02
- 0.04~dex, with the only exception being that HD\,117176 now reveals 
Ba abundance that is higher by 0.06~dex. The latter star shows the largest
discrepancy in $\eps{Ba}$ derived from the $\lambda\,5853$ and
$\lambda\,6496$ lines.

\begin{figure}
\resizebox{88mm}{!}{\includegraphics{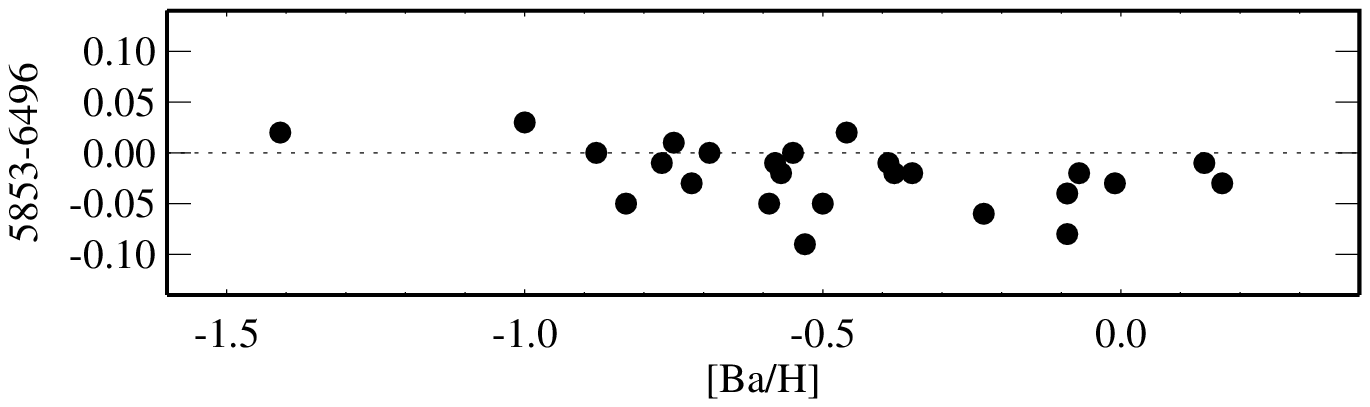}}
\vspace{-0mm}
\resizebox{88mm}{!}{\includegraphics{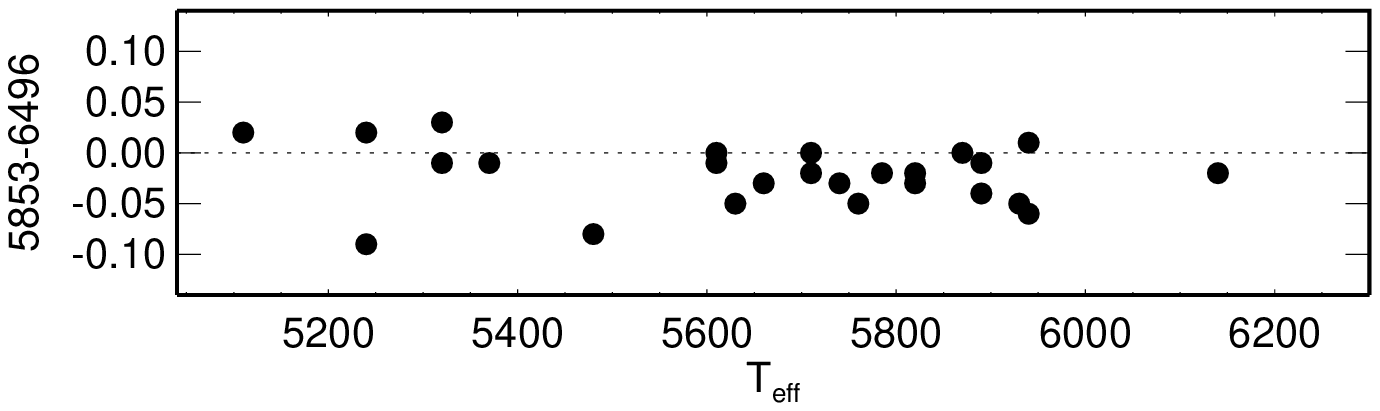}}
\vspace{-0mm} \caption[]{NLTE abundance differences
$\eps{}(\lambda5853) - \eps{}(\lambda6496)$ versus total Ba
abundance [Ba/H] (top panel) and $\Teff$ (bottom panel) }
\label{5853_6496}
\end{figure}

Ba abundance determined from the $\lambda\,5853$ line turns 
out to be systematically lower than that from the second line; 
the Ba abundance is lower by
$-0.03\pm0.02$~dex in the stars with [Ba/H] $\ge$ --0.35, by
$-0.01\pm0.03$~dex in the more Ba-poor stars, and by
$-0.02\pm0.03$~dex for the whole sample (Fig.~\ref{5853_6496}, top
panel). The abundance difference $\eps{}(\lambda5853) -
\eps{}(\lambda6496)$ shows no correlation with $\Teff$
(Fig.~\ref{5853_6496}, bottom panel). 
 Ba abundance from the $\lambda\,5853$ line can be underestimated due to 
overestimation of the oscillator strength of the blending Fe~I $\lambda5853$ 
line. Using $\log gf$ = --2.371 calculated recently by Kurucz 
(http://cfaku5.cfa.harvard.edu), we derive the larger Ba abundance from the 
$\lambda\,5853$ blend: it is larger by 0.01~dex in HD\,9407 ([Ba/H] = --0.01) and by 
0.02~dex in HD\,68017 ([Ba/H] = --0.50) and HD\,64606 ([Ba/H] = --1.00). 
Another source of the obtained discrepancy between the two \ion{Ba}{ii} lines 
is the uncertainty of the van der Waals damping constant of 0.05 -- 0.18 dex according to Barklem (\cite{barklem06}).
In the 
next subsection, we will discuss an effect of total Ba abundance error on the 
Ba even-to-odd isotope ratio in stars. 


%
%


\subsection{Ba even-to-odd isotope ratio in stars}

The fraction of the odd isotopes of Ba is derived from the
requirement that Ba abundances derived from the \ion{Ba}{ii}\,
resonance line and the subordinate lines must be equal.  We emphasize that 
absolute Ba abundance $\eps{Ba}$ is used in this procedure, and, thus, the 
obtained fraction does not depend on the adopted value of solar Ba abundance. 

\begin{figure}
\resizebox{88mm}{!}{\includegraphics{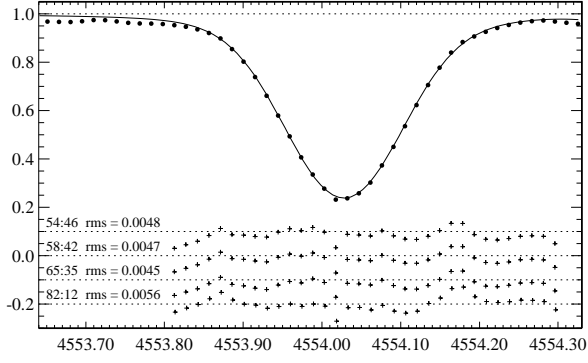}}
\vspace{-3mm} \caption[]{Synthetic NLTE flux profile corresponding
to $\eps{Ba}$ = 1.52 and Ba even-to-odd isotope ratio 58 : 42
(continuous line) compared to the observed profile (bold dots) of the
Ba~II resonance line in HD\,10519. $\eps{Ba}$ = 1.51 for the isotope ratio 54
: 46, $\eps{Ba}$ = 1.54 for 65 : 35, and $\eps{Ba}$ = 1.64 for 82 : 18.
The (O - C) values multiplied by a factor of 5 are shown for
various isotope mixtures in the lower part of the figure. See text for
more details} \label{star_4554}
\end{figure}

Figure \ref{star_4554} illustrates a determination procedure for one
of the stars of our sample, HD\,10519. The same Ba abundance,
$\eps{Ba}$ = 1.52, is obtained from both subordinate lines. The
$\lambda\,4554$ line gives $\eps{Ba}$ = 1.64 at the assumption of
the isotope mixture in solar system matter and $\eps{Ba}$ = 1.51 for
the pure $r-$process yields predicted by Arlandini et al.
(\cite{rs99}). As noted above, the synthetic $\lambda\,4554$ line
profile depends on the total fractional abundance of the
odd isotopes rather than on the exact abundance of each odd isotope.
Therefore, we vary the even-to-odd isotope ratio (from here on, 
``the isotope ratio'') in calculations of $\lambda\,4554$
until the obtained Ba abundance agrees with that from the
subordinate lines. The required Ba abundance $\eps{Ba}$ = 1.52 is achieved from
$\lambda\,4554$ when the isotope ratio equals 58 : 42. The following 
points should be noted.
\begin{itemize}
\item When we go from the isotope ratio 82 : 18 to 54 : 46, the
relative shift of the observed profile by 0.001~\AA\, is required
to achieve the best fit. \item We define the best fit from a
requirement of the minimum $rms$ difference between observed and
calculated spectra. The equivalent width corresponding to the best
fit may be different for different isotope ratios, e.g., in our
calculations for HD\,10519, the equivalent width measured over the
same spectral interval, 4553.8\AA\,- 4554.3\AA, W($\lambda\,4554$)
= 154.1~m\AA\, for the isotope ratio 82 : 18 and 152.2~m\AA\, for
54 : 46. This can be understood. The smaller fraction
of the odd isotopes that the larger Ba abundance is required to fit, in the
observed line profile, and the calculated line wings, \emph{which
never perfectly fit the observed ones} (Fig.~\ref{star_4554}),
contribute to W$_\lambda$ in the greater extension, when compared to the
case of the larger fraction of the odd isotopes with the smaller Ba
abundance. \item The macroturbulence value was allowed to be free
because we have no arguments to fix it. An increase of
the fraction of the odd isotopes cancels, in part, a saturation of
$\lambda\,4554$, and the line formation depth shifts downward; e.g.,
in our calculations for HD\,10519, a macroturbulence value grows
from 4.25\kms\, to 4.5\kms\, when the isotope ratio changes from 
82 : 18 to 54 : 46.
\end{itemize}

{\it Uncertainty of the fractional abundance of the odd isotopes
of Ba in stars.} 
Random errors of a desirable fraction are due to
total Ba abundance error and the uncertainty of stellar
parameters, $\Teff$, $\log g$, and microturbulence value. Systematical
errors are caused by the uncertainty of the \ion{Ba}{ii}
$\lambda\,4554$ atomic parameters, namely, $f_{ij}$ and $C_6$ values. 
Varying  all mentioned 
parameters, we have performed the test calculations for
the three representative stars of our sample with different values
of [Ba/H], HD\,9407 ([Ba/H] = -0.01), HD\,68017 ([Ba/H] = -0.50),
and HD\,64606 ([Ba/H] = -1.00). The effect on the Ba odd isotope
fraction in these stars is shown in Table~\ref{uncertain}.

The most important source of total Ba abundance error in the stars
with metallicity close to solar one is the uncertainty of the van der
Waals damping constant of the subordinate lines. A variation in
$\log C_6$ of 0.1 translates to the 0.03~dex variation in
$\eps{Ba}$. When $\lambda\,4554$ is saturated, a reduction of Ba
abundance obtained from the subordinate lines by 0.01~dex requires
an increase of the fractional abundance of the odd isotopes by
$\sim$2\%\, to $\sim$3\%\, depending on the resonance line strength.
A variation of 80~K in $\Teff$ 
translates to 
the uncertainty of the obtained fraction of 3\%\, to 4\%. In the stars with 
[Ba/H] $\le -0.5$, a variation in $\Vmic$ has nearly the same effect on 
both the resonance and subordinate lines, 
and no change of the Ba odd isotope fraction is required. In the
stars with metallicity close to solar one, the resonance line lies
on the damping part of the curve of growth and is weakly sensitive
to $\Vmic$ variations, while this is not the case for the subordinate
lines. The uncertainties of total Ba abundance, $\Teff$, $\log g$, and 
$\Vmic$, in total, result in a random error of the obtained fraction 
of 7\%, 9\%, and 10\%\, in the stars with [Ba/H] = 0, --0.5, 
and $-1$, correspondingly.
 There can be also a systematical
difference of 4\%\, to 5\%\, between the thick disk and thin disk stars,
due to a discrepancy in the mean value $\eps{Ba}(\lambda5853) -
\eps{Ba}(\lambda6496)$ between the samples of thick disk and thin disk
stars.

\begin{table}[htbp]
\caption{Effect on the Ba odd isotope fraction (in \%) caused by
uncertainties of atomic data and stellar parameters }
\label{uncertain}
\begin{center}
\begin{tabular}{llccc}
\hline
Input & Input & \multicolumn{3}{c}{[Ba/H]} \\
\cline{3-5}
parameter & error & --0.01 &  --0.50 & --1.00 \\
\hline
 \ $\log C_6 (5d - 6p)$ & +0.1       & +6    & +3.8 & +1.4 \\
 \ $\eps{Ba}$           & --0.02     & +4    & +5   & +5.6 \\
 \ $\Teff$(K)           & +80        & +3    & +4   & +4   \\
 \ $\log g$             & --0.1      & $< 1$ & --1  & +3  \\
 \ $\Vmic$(\kms)        & +0.1       & +5    & $< 1$ & $< 1$ \\
 \ $\log gf (\lambda4554)$ & --0.02  & +4    & +5   & +5.6 \\
 \ $\log C_6 (\lambda4554)$ & +0.1   & --7   & --8  & --8 \\
\hline
\end{tabular}
\end{center}
\end{table}

A variation of 0.1~dex in the van der Waals damping constant of
the $\lambda\,4554$ line produces a strong effect, 7~\%\, to 8~\%\, of
the derived fraction of the odd isotopes of Ba; however, this is a
systematical effect. It depends only weakly on Ba/H and therefore
will not affect a difference in the fraction of the odd isotopes
between different stars.

We suppose that in each individual star, the spread of Ba abundance
derived from two subordinate lines fully reflects both random
and systematical errors of total Ba abundance, including effects of
the uncertainties of atomic parameters ($gf$ and $C_6$ values),
microturbulence value, and NLTE treatment. It is worth while to remember
that due to different strengths of the \ion{Ba}{ii}
$\lambda\,6496$ and $\lambda\,5853$ lines ($f(\lambda6496) /
f(\lambda5853) \simeq 4$), they lie  in each star on different
parts of the curve of growth. The uncertainty of a desirable
fraction in each star is evaluated, taking into account its
individual Ba abundance error and uncertainty of the
$\lambda\,4554$ line profile fitting. In terms of Ba abundance, the
latter value is estimated to be 0.01~dex, based on the analysis of the (O
- C) values. For example, in HD\,10519, both subordinate lines give
the same $\eps{Ba}$ and the lower and upper limits of the odd
isotope fraction are calculated taking into account only the
uncertainty of the $\lambda\,4554$ line profile fitting
(0.01~dex). In the range of the odd isotope fraction between
35\%\, and 46\%, characteristic of this star, a Ba abundance
variation of 0.01~dex produces the 4~\%\, variation in the
fractional abundance of the odd isotopes. Therefore, our estimate
of the fraction of the odd isotopes of Ba in this star is 42 $\pm$
4\%.

The obtained fraction of the odd isotopes, with its lower and
upper limits in the stars of our sample, are presented in 
Table~\ref{startab}. 

\section{Stellar Ba odd isotopes as an indicator of the $r/s-$process
nucleosynthesis}\label{concl}

\begin{figure}
\resizebox{88mm}{!}{\includegraphics{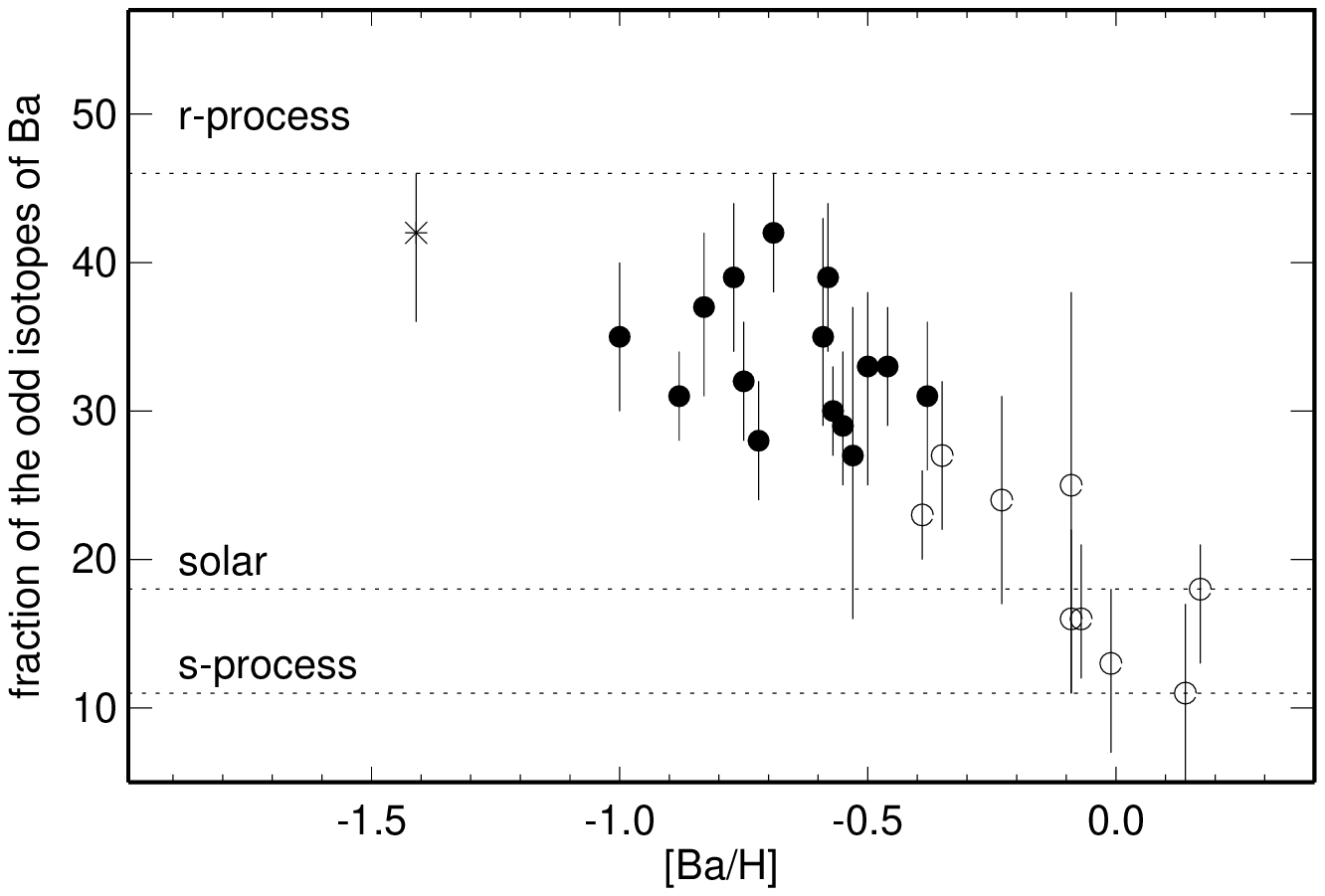}} \vspace{-0mm}
\resizebox{88mm}{!}{\includegraphics{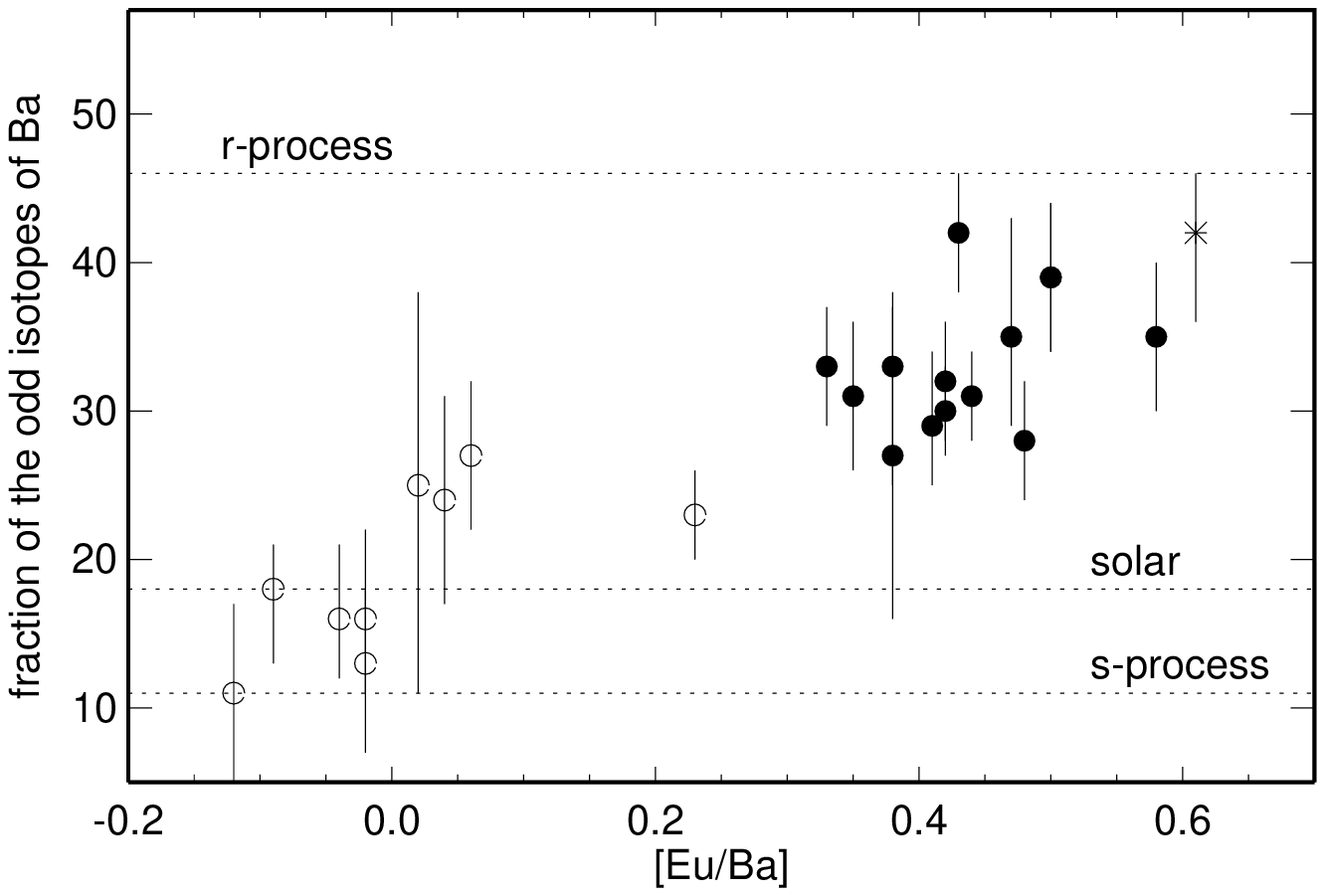}}
\vspace{-0mm} \resizebox{88mm}{!}{\includegraphics{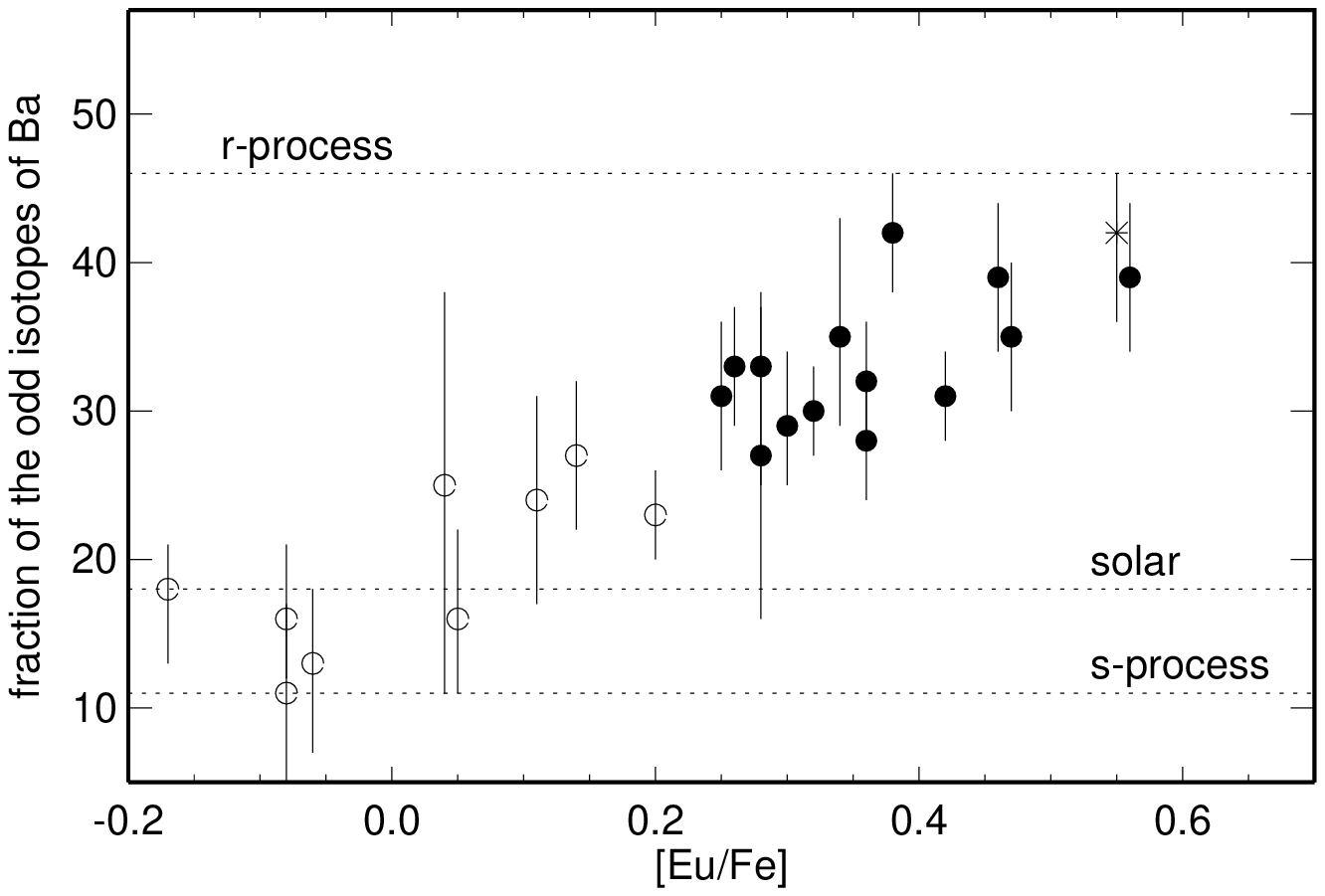}}
\vspace{-3mm} \caption[]{The fractional abundance of the odd
isotopes of Ba versus [Ba/H] (top panel), [Eu/Ba] (middle panel),
and [Eu/Fe] (bottom panel). The uncertainty of individual values
is shown by vertical lines. Horizontal lines indicate the solar
fraction of the odd isotopes, 18\%, and the values predicted by
Arlandini et al. (1999) and Travaglio et al. (1999) for a pure
$r-$process, 46\%, and pure $s-$process, 11\%, production of
barium} \label{ba_iso}
\end{figure}

In Fig. \ref{ba_iso}, we plot the fractional abundance of the odd
isotopes of Ba versus the total Ba abundance and the [Eu/Ba] and [Eu/Fe]
abundance ratios. Based on the spread of Ba abundance in each star
(Fig.~\ref{5853_6496}), error bars of the [Ba/H] value do not
exceed 0.1~dex. The [Eu/Fe] abundance ratios are taken from our
earlier determinations (Mashonkina \& Gehren \cite{eubasr}), where
their errors are estimated as $\Delta$[Eu/Fe] = 0.05~dex. We
note that both Eu and Fe abundances are derived from weak spectral
lines of the dominant ionization stage, \ion{Eu}{ii} and \ion{Fe}{ii},
respectively, and the uncertainty of stellar parameters has a 
negligible effect on the derived Eu/Fe abundance ratio. Error bars
of the [Eu/Ba] value are mainly defined by errors of Ba abundance.

It is clearly seen that the obtained fraction grows towards the
lower Ba abundance. Its value in the thick disk stars 
ranges between 27\%\, and 42\%, with the mean value 33$\pm$4\%. This
indicates the higher contribution of the $r-$process to barium in
the thick disk stars compared to the solar system matter in
agreement with our expectations. Analysis of the [$\alpha$/Fe]
abundance ratios (we cite here only the first such studies:
Gratton et al. \cite{Gratton96}, \cite{Gratton00}; Fuhrmann
\cite{Fuhr3}; Prochaska et al. \cite{proch}) suggests that the
thick disk stellar population is nearly the same age as the
Galactic halo one. This conclusion was confirmed from an analysis of
Eu/Ba and Nd/Ba abundance ratios (Mashonkina \& Gehren
\cite{euba}; Mashonkina et al. \cite{mash_nd}; Bensby et al.
\cite{bensby05}). The significant overabundance of Eu and Nd relative
to Ba found in the thick disk stars at [Fe/H] $\le -0.3$ says that
these stars formed in the early Galaxy when SNeII dominated the 
synthesis of heavy elements.

Both the [Eu/Ba] abundance ratio and the fractional abundance of
the odd isotopes of Ba are sensitive to the relative contribution of
the $r-$process to heavy element synthesis, and it is natural to
expect their correlation. Figure~\ref{ba_iso} shows that such
correlation really exists. It is more clearly seen in the plane
{\it fractional abundance of the odd isotopes of Ba} - [Eu/Fe]
because the range of [Eu/Fe] values in the thick disk stars is
larger than that of [Eu/Ba], and errors of the [Eu/Fe]
values are smaller than that of [Eu/Ba].

 The obtained data provide observational constraints to models of the $s-$
and $r-$process nucleosynthesis. The "classical" model of the $s-$process based
on the newest accurate measurements of the stellar $n-$capture
cross-sections (Arlandini et al. \cite{rs99}) predicts a pure $s-$process 
origin of solar even isotopes of Ba. In this model, no even isotope of Ba 
appears in stars formed in the early Galaxy before the onset of the 
$s-$process in AGB stars. In contrast to this prediction, 
we find a significant fraction of the 
even isotopes of Ba, $\sim67$\%, in old Galactic stars, the thick disk
stars. The "stellar" models of Arlandini et al. (\cite{rs99}), based on 
stellar AGB models of 1.5 $M_\odot$ and 3 $M_\odot$ with half solar
metal abundance, and Travaglio et al. (\cite{eu99}), who integrate 
$s-$abundances from different generations of AGB stars, i.e., consider the whole
range of Galactic metallicities, both predict that solar abundance of isotope 
\iso{138}{Ba} is contributed to by the $s-$process only in part: 86\%\, and 84\%\,, respectively. As a result, in the oldest stars of the Galaxy, formed 
at the epoch of SNeII dominance in nucleosynthesis, a fraction of the even 
isotopes 
of Ba is expected to be at the level of 54\%. Ba isotopic fractions found in the 
thick disk stars favor the "stellar" model of heavy element synthesis.

\begin{acknowledgements}
We are very grateful to Klaus Fuhrmann for providing reduced
stellar spectra and Thomas Gehren for providing a Windows version of
the code DETAIL. ML acknowledges with gratitude the National
Astronomical Observatories of Chinese Academy of Science for 
warm hospitality during a productive stay in October - December
of 2005. We thank the anonymous referee for useful remarks and comments. 
This research was supported by the Deutsche
Forschungsgemeinschaft with grant 436 RUS 17, the Russian
Foundation for Basic Research with grant 05-02-39005-GFEN-a, the
Natural Science Foundation of China with grants NSFC 10433010 and 10521001, the
RF President with a grant on Leading Scientific Schools 1789.2003.2, and 
the Presidium RAS Programme ``Origin and evolution of stars and the Galaxy''.
\end{acknowledgements}

\appendix
\section{The outlook for a determination of Ba isotopic fractions in very 
metal-poor stars}
It would be very important to extend our study 
to the older and more metal-poor stars revealing a pure $r-$process 
nucleosynthesis and to determine from observations the relative 
yields of the even and odd isotopes of Ba in the $r-$process. 
At [Ba/H] $< -2$, our method fails to
give a reliable value of the fractional abundance of the odd
isotopes of Ba due to (i) less sensitivity of the $\lambda\,4554$
line to its variation, and (ii) less accuracy of total Ba
abundance. An increase of the fraction of the odd isotopes of Ba
from 18\%\, to 46\%\, leads to a decrease of Ba abundance
derived from $\lambda\,4554$ by 0.07~dex at [Ba/H] = --2 and by
only 0.02~dex at [Ba/H] = --3.3. At the same time, the weaker
subordinate line $\lambda\,5853$ becomes, in fact, unmeasurable in
stellar spectra.

Can we determine a desirable fraction based on only the HFS
broadening of the $\lambda\,4554$ line without knowing total Ba
abundance? Let's imagine that we have a perfect observed profile of
$\lambda\,4554$. Then we fit it varying the fractional
abundance of the odd isotopes of Ba. We have modelled such
situation for two "stars" using two model atmospheres.

"Star" 1: $\Teff =$ 5710 K, $\log g =$ 4.00, [M/H] = --0.64,
$\Vmic =$ 1.1~\kms, [Ba/H] = --0.69. The theoretical line profile
of $\lambda\,4554$ calculated assuming the Ba odd isotope
fraction of 42\%\, and convolved with the Gaussian of
4~\kms\, and radial-tangential profile of $\Vmac =$ 4.5~\kms 
will serve as the "observed" line profile. Its W$_\lambda$ = 152 m\AA.

"Star" 2:  $\Teff =$ 6350 K, $\log g =$ 4.03, [M/H] = --2.07,
$\Vmic =$ 1.7~\kms, [Ba/H] = --2.14. The "observed" profile of
$\lambda\,4554$ was obtained assuming the Ba odd isotope fraction
of 46\%\, and applying the Gaussian of 2~\kms\, and $\Vmac =$ 4.7~\kms. Its
W$_\lambda$ = 50 m\AA.

\begin{figure}
\resizebox{88mm}{!}{\includegraphics{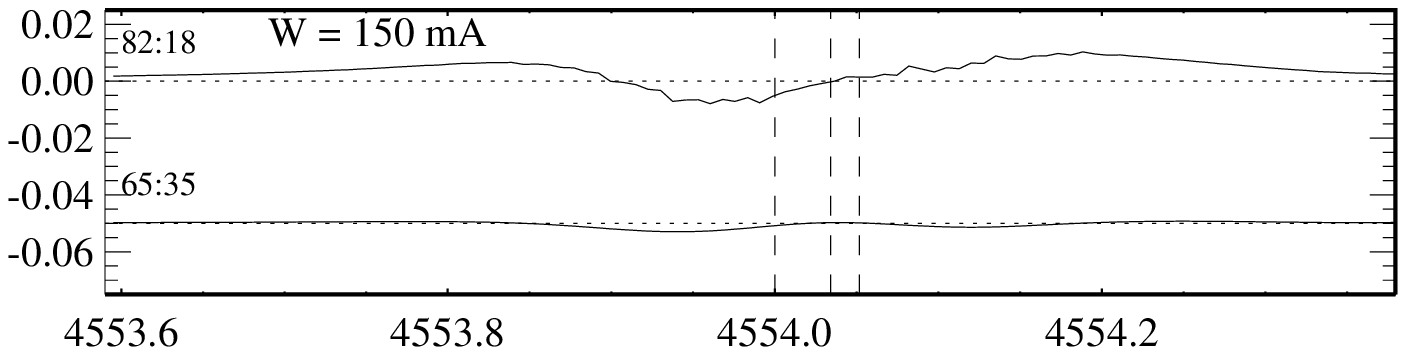}}
\vspace{-0mm}
\resizebox{88mm}{!}{\includegraphics{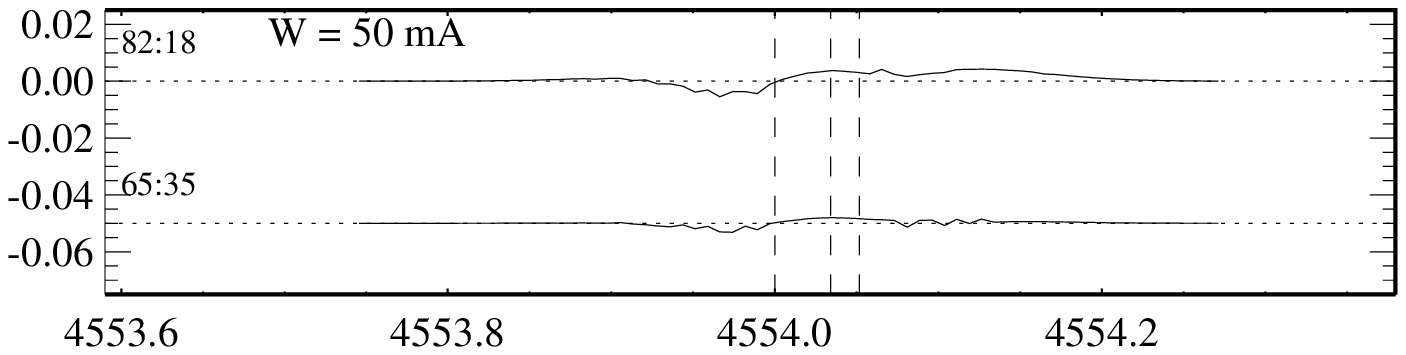}}
\vspace{-3mm} \caption[]{The differences between "observed" and
calculated spectra, ("O" - C) in "star" 1 (top panel) and
"star" 2 (bottom panel) for the even-to-odd isotope ratios 82 : 18 and 
65 : 35. In the latter case, the differences shifted down by 0.05 are shown. 
The HFS components of the odd isotopes appear in two groups shifted
relative to the resonance line of the even isotopes at 4554.034
\AA\, by 18 m\AA\, and $-34$ m\AA. Their positions are indicated
by vertical lines. See text for more details} \label{4554th}
\end{figure}

Assuming the fraction of the odd isotope of Ba, $odd$ = 18\%\, and then 
35\%, we obtained the best fits for each case and for both "stars".
Ba abundance and $\Vmac$ value were allowed to vary. The Gaussian was
fixed. In practice, the $\Vmac$ values turn out to be within the 
uncertainty of determination: $\Vmac$ = 4.25 \kms\, when $odd$ =
18\%\, and 4.4 \kms\, when $odd$ = 35\%\,  in "star" 1;
$\Vmac$ = 5.1 \kms\, when $odd$ = 18\%\, and 5.0 \kms\, when $odd$
= 35\%\, in "star" 2. 
The ("O" - C) differences are shown in Fig.~\ref{4554th}.

It can be seen from Fig.~\ref{4554th} that an accuracy of line
profile measurements has to be much higher than 0.05\%\, if one
wishes to determine a desirable fraction within the uncertainty of
10\%.

{\bf Note added in proofs.} 

For $\lambda\,4554$, the van der Waals damping constant based on theoretical predictions of Barklem \& O'Mara (\cite{omara}) is equal to $\log C_6$ = --31.46 with the uncertainty of 0.05 -- 0.18 dex according to Barklem \& Aspelund-Johansson (\cite{baj05}) and Barklem (\cite{barklem06}). Using this value and the MAFAGS solar model atmosphere, we find solar Ba abundance $\eps{\odot Ba} = 2.17$, in well agreement with the recent determinations of 
the meteoritic ($\eps{met,Ba} = 2.16\pm0.03$) 
and solar ($\eps{\odot Ba} = 2.17\pm0.07$) Ba abundance by Asplund et al.
(\cite{met05}). Assuming $\log C_6$ = --31.46 for $\lambda\,4554$, we checked  other stars of our sample with stellar parameters close to solar ones, HD\,9407, HD\,10697, HD\,134987, and HD\,168009 (see Sect.\,\ref{obs}), and found systematically lower Ba abundance from the \ion{Ba}{ii} resonance line compared to that from the subordinate lines, by 0.06 dex to 0.10 dex. Variation of stellar parameters and microturbulence value does not help to achieve agreement between different lines, and the only way out is to reduce collisional broadening of the resonance line. Therefore, we apply in this study the value $\log C_6$ = --31.65 to $\lambda\,4554$. For the \ion{Ba}{ii} multiplet $5d - 6p$, the van der Waals damping constant $\log C_6 = -31.3$ found empirically in Sect.\,\ref{data} is very close to $\log C_6 = -31.28$ based on the theory of Barklem \& O'Mara (\cite{omara}) and accessible via the Vienna Atomic Line 
Data base (VALD, Kupka et al. \cite{vald}.

\end{document}